\documentclass[final,1p,times]{elsarticle}

\usepackage{graphicx}
\usepackage{amssymb}
\usepackage{amsthm}
\usepackage{lineno}

\journal{Nuclear Physics A}

\begin{document}

\begin{frontmatter}

\title{Two particle correlation measurements with respect to higher harmonic event planes at PHENIX}

\author{Takahito Todoroki (for the PHENIX\fnref{col1} Collaboration)}
\fntext[col1] {A list of members of the PHENIX Collaboration and acknowledgements can be found at the end of this issue.}
\address{University of Tsukuba,Tsukuba, Ibaraki 305, Japan\\
 RIKEN Nishina Center for Accelerator-Based Science, Wako, Saitama 351-0198, Japan}


\begin{abstract}
Measurements of two particle azimuthal correlations in relativistic heavy ion collisions provide
information of the possible interplay between hard-scattered partons and the hot-dense medium.
Toward an understanding of parton-medium coupling, it is indispensable to obtain correlations
where contributions from higher harmonic flow($v_n$) are rejected. It is also important to produce correlation measurements where the trigger particle is selected relative to second and third order event planes. This enables us to explore path-length dependence of parton energy loss and the influence of the medium on the jets. We present the latest PHENIX results of correlations in which contributions from higher harmonic flow have been subtracted, as well as second and third order event plane-dependent correlations in Au+Au collisions at $\sqrt{s_{NN}}=200$\,GeV.
\end{abstract}

\end{frontmatter} 


\section{Introduction}
In relativistic heavy ion collisions, the triangularity, squarity (fourth-order), and higher-order deformations originate from fluctuations in the initial participant geometry. This gives rise to higher-order flow harmonics where the anisotropy of particle productions, relative to each harmonic event plane, is observed over a wide rapidity range.
Their existence has been revealed by studies with the AMPT model\cite{Alver:2010prc81}
and experimental measurements at the RHIC\cite{Adare:2011prl107} and the LHC\cite{Aamodt:2011prl107}\cite{Aad:2012prc86}.

The amplitude of flow harmonics $v_n$ is defined as
\begin{eqnarray}
\quad\quad
\frac{dN}{d\phi}
\propto
1 + \sum_{n=1} 2 v_{n} \cos (n(\phi-\Psi_{n})),\quad
v_n = \left< \cos ( n (\phi - \Psi_{n}) ) \right> , (n=1,2,3,...),
\label{eq:vn}
\end{eqnarray}
where $\phi$ is the azimuthal angle of particle and $\Psi_n$ is the direction of the $n^{th}$ order event plane in the transverse plane. The $v_n$ of hadrons within $|\eta|<0.35$ is measured with event planes determined
in $1.0<\eta<2.8$ to ensure a sufficient pseudo-rapidity gap between particle and event planes to reduce
non-flow correlations due to jet production.

Two particle correlations are calculated by dividing the relative angular distributions of real event pairs
with those from mixed events, and applying proper normalization:
\begin{eqnarray}
C(\Delta\phi)=N^{real}_{pair}(\Delta\phi)/N^{mix}_{pair}(\Delta\phi)\int d\Delta\phi N^{mix}_{pair}(\Delta\phi)/\int d\Delta\phi N^{real}_{pair}(\Delta\phi),
\label{eq:two-particle}
\end{eqnarray}
where $\Delta\phi=\phi^{asso}-\phi^{trig}$ is the relative angle between trigger and associated particles.
The correlations are measured using particles emitted in the range $|\eta|<0.35$, i.e. without rapidity gaps between trigger and associate particles. In these correlations, the contributions from jets survive,
thus allowing us to study the interplay between hard-scattered partons and the medium.
The expected flow shapes included in these correlations can be estimated from experimentally measured
event plane resolutions and $v_n$, and are subtracted with the normalization factor determined by the ZYAM method\cite{Adler:2006prl97}.

\section{Two particle correlations with $v_n$ contribution subtractions}
\begin{figure}[!t]
\centering
\includegraphics[width=1.0\textwidth]{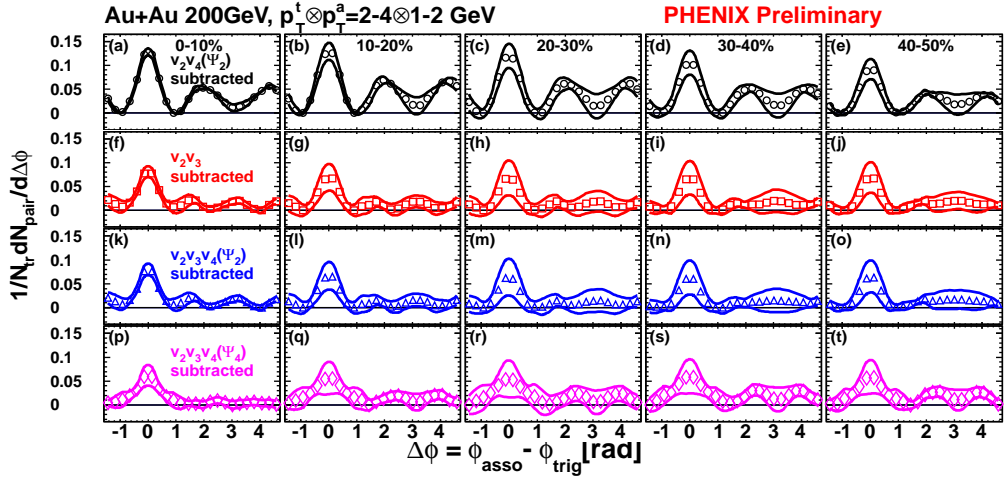}
\caption{
(Color online)
Two particle charged hadron correlations in Au+Au collisions at $\sqrt{s_{NN}}=200$ GeV
in five centrality intervals in the range 0\%-50\%.
The transverse momentum ranges are 2$<$$p_T$$<$4\,GeV/$c$ and 1$<$$p_T$$<$2\,GeV/$c$ for trigger and associates respectively.
Contributions of $v_2$, $v_3$, $v_4(\Psi_2)$ and $v_4(\Psi_4)$ are subtracted using the ZYAM method.
The solid lines are the systematic uncertainties propagated from higher-order flow harmonics.
}
\label{fig:1}
\end{figure}
Figure\ref{fig:1} shows the two particle charged hadron correlations in Au+Au collisions
at $\sqrt{s_{NN}}=200$\,GeV in five centrality intervals in the range 0\%-50\%.
The trigger particle is required to have a transverse momentum of 2$<$$p_T$$<$4\,GeV/$c$ which are correlated with associates in the range 1$<$$p_T$$<$2\,GeV/$c$.
Contributions from various combinations of $v_2$, $v_3$, $v_4(\Psi_2)$ and $v_4(\Psi_4)$ are subtracted using the ZYAM method.

The away side double hump structure\cite{Adare:2008prc77} is seen in the correlations with subtraction of $v_2$ and $v_4(\Psi_2)$
contributions in all centrality intervals, but it is largely reduced by subtracting the contributions of $v_2$ and $v_3$.
The inclusion of $v_4(\Psi_2)$ to the subtraction of $v_n$ contributions does not change the away side residuals
of the $v_2$ and $v_3$ subtracted correlations. However, $v_4(\Psi_4)$ does disturb the away side correlations.
While the away side double hump is almost removed in central(0-10\%) collisions, it reappears in mid central(20-50\%) collisions.
The treatment of $v_4$ in addition to $v_3$ appears to be most important for determining whether displaced peaks may appear in the subtracted results.
There are however currently uncertainties related to further effects of correlations between the different-order event planes which are still under investigation before strong conclusions can be made.

\section{Two particle correlations versus second and third order event planes}
\begin{figure}[!th]
\centering
\includegraphics[width=1.0\textwidth]{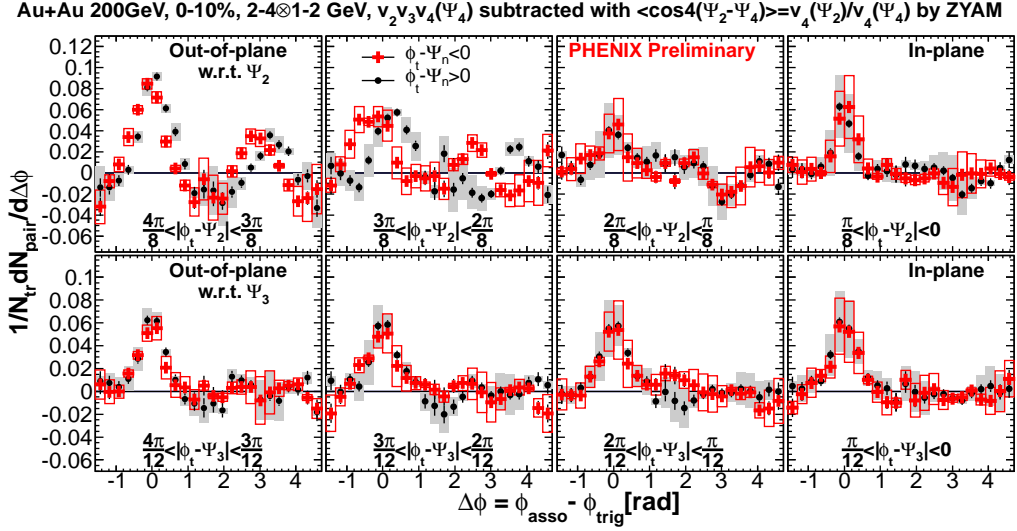}
\caption{
(Color online)
Two particle charged hadron correlations with trigger particle selection relative to second(top panels)
and third(bottom panels) order event planes in Au+Au collisions at $\sqrt{s_{NN}}=$200 GeV in most-central(0-10\%) collisions.
The transverse momentum ranges are 2$<$$p_T$$<$4\,GeV/$c$ and 1$<$$p_T$$<$2\,GeV/$c$ for trigger and associates respectively.
The boxes are the systematic uncertainties propagated from higher-order flow harmonics.
}
\label{fig:2}
\end{figure}

\begin{figure}[!t]
\centering
\includegraphics[width=1.0\textwidth]{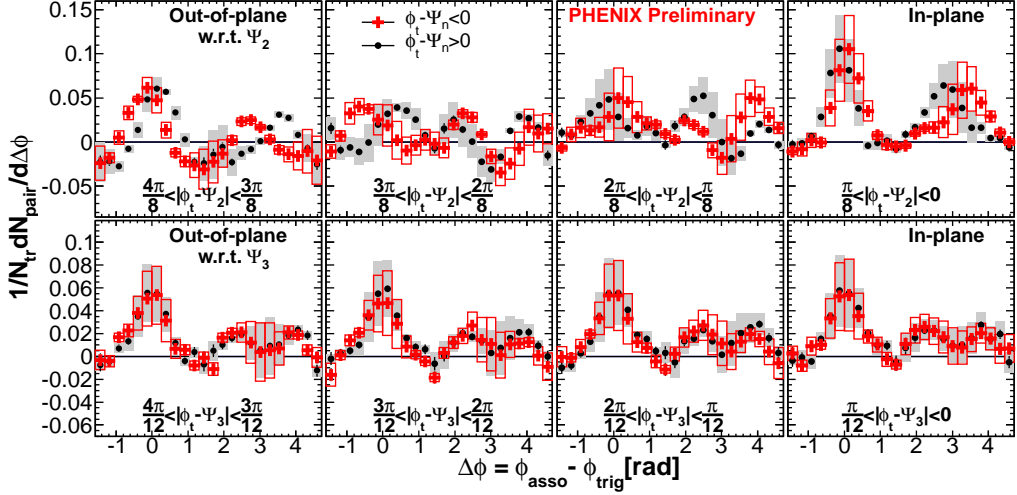}
\caption{
(Color online)
Same two particle charged hadron correlations as Fig.2 but for mid-central(40-50\%) collisions.
}
\label{fig:3}
\end{figure}

Figure\ref{fig:2} and \ref{fig:3} show the two-particle charged hadron correlations with trigger particle selection relative to the second and third order event planes in Au+Au collisions at $\sqrt{s_{NN}}=$200 GeV in most-central(0-10\%) and mid-central(40-50\%) collisions.
The transverse momentum ranges are 2$<$$p_T$$<$4\,GeV/$c$ and 1$<$$p_T$$<$2\,GeV/$c$ for trigger and associates respectively.
Contributions from $v_2$, $v_3$ and $v_4(\Psi_4)$ are subtracted from each triggered correlation
by means of the ZYAM method with a single contribution level determined in correlations without trigger selection, for all event plane bins.

In this analysis, the trigger particle is selected not only depending on the relative angle between the trigger
particle and the event planes but also on left or right side of the event planes, separately.
The left (right) side of the event planes is defined as $\phi^{trig}$$-$$\Psi_n$$<$0 ($\phi^{trig}$$-$$\Psi_n$$>$0).
It is expected that correlations show a left-right asymmetry if there is parton-medium coupling,
since the paths of partons in the medium have a mirror like asymmetry between
the cases where the trigger particle is selected in the left or right side of the event planes.
The black (red) points represent the correlations where the trigger is selected in the left (right) side of the event planes.
The second order event plane dependent correlations show a left-right asymmetry
which is more pronounced in mid-central collisions than in most-central collisions.
The is consistent with the expected average eccentricity difference therein.  The correlations also show the dependence on relative angle between the trigger particle and the second order event plane. The yields of correlations with trigger particle in the in-plane direction are more pronounced compared to correlations with trigger particle in the out-of-plane direction in mid-central collisions, and left/right asymmetry is clear in both directions. This is consistent with path length dependence of parton energy loss, although there is an opposite trend in central events.

In the third order event plane dependent correlations, there is no left-right asymmetry and no trigger angle dependence observed in either centrality class.  All the correlations with trigger particle selection relative to the event plane are consistent within systematic uncertainties.  If further uncertainties can be resolved concerning the subtraction (see above) these various features could be indications of energy loss effects and/or non-trivial jet-medium interplay, including a lack thereof for the 3rd order plane (to within the PHENIX event plane resolutions.)

\section{Summary}
The latest PHENIX results of correlations with the subtraction of contributions from higher harmonic flow as well as second and third order event plane dependent correlations in Au+Au collisions at $\sqrt{s_{NN}}=200$ GeV have been presented.
The treatment of both $v_4$ in addition to $v_3$ are crucial in determining
the away side residual shape of the correlations. The second order event plane dependent correlations
show the clear left-right asymmetry and the dependence on relative angle between trigger particle
and event plane. The third order event plane dependent correlations do not show the such behavior.  The correlations indicate different sensitivity the second and third order event planes. These observations could have implications related to energy-loss and/or jet-soft particle interplay.

\end{document}